\documentclass{PoS}

\title{Vector meson electromagnetic form factors}

\ShortTitle{Vector meson electromagnetic form factors}

\author{\speaker{B.~G.~Lasscock}, J.~Hedditch\thanks{Project Leader} , D.~B.~Leinweber,\
        A.~G.~Williams\\        
        Special Research Centre for the
        Subatomic Structure of Matter,
        and Department of Physics,
        University of Adelaide, Adelaide SA 5005,
        Australia \\
        E-mail: \email{blasscoc@physics.adelaide.edu.au}}

      \abstract{ 
                 The charge, magnetic and quadrupole form factors of vector mesons and the 
                 charge form factor of pseudo-scalar mesons are calculated in quenched 
                 lattice QCD. The charge radii and magnetic moments are derived. 
                 The quark sector contributions to the form factors are calculated separately
                 and we highlight the environmental sensitivity of the light-quark contribution 
                 to charge radii. 
               }

\FullConference{XXIV International Symposium on Lattice Field Theory\\
		 July 23-28 2006\\
		 Tucson Arizona, US}

\begin{document}

\section{Introduction}

We calculate the charge, magnetic and quadrupole form factors of
vector mesons and the charge form factor of pseudo-scalar mesons
in quenched lattice QCD.  In each case the charge radii and magnetic
moments are derived.  

Our aim is to study to what extent the qualitative quark model picture
is consistent with quenched lattice QCD. Interestingly, it has been
shown in a lattice calculation by Alexandrou et~al.\
\cite{Alexandrou:2002nn} that the distribution of charge in the vector
meson is oblate, and therefore not consistent with the picture of a
quark anti-quark in relative S-wave.  By calculating the vector meson
quadrupole form factor we make a direct comparison with the findings
of Ref.~\cite{Alexandrou:2002nn}.

For each observable we calculate the quark sector contributions
separately. Using this additional information we examine the
environmental sensitivity of the light-quark contributions to the
pseudo-scalar and vector meson charge radii and we can measure the
dominance of the light quark contributions to the $K$ and $K^*$.

\subsection{ 2-pt Correlation Function }

We begin the discussion with a brief description of how we extract the
meson two-point functions on the lattice, followed by a review of the
extraction of observables from the meson 3-pt functions. 

The vector meson 2-pt correlation function is defined by,
\begin{eqnarray*}
\label{eq:2pt}
G^{\mu\nu}(t,\vec{p}) &=& \sum_{\vec{x}}e^{-i\vec{p}\cdot \vec{x}}\
\langle\Omega| \chi^\mu (x)   {\chi^\nu}^{\dagger} (0) |\Omega\rangle\ .
\end{eqnarray*} 
where in this case $\chi^{\mu}$ is the standard $\rho^{+}$-meson
interpolating field $\chi^{\mu} = \bar{d}^{a}\gamma^{\mu}u^{a}$
\cite{Hedditch:2005zf}, here $u,d$ is the quark flavour, and $a$ is a
colour label.  To evaluate this function we first insert a complete
set of energy, spin and momentum states,
\begin{eqnarray*}
G^{\mu\nu}(t,\vec{p}) &=& \sum_{s} e^{-E_{\rho}t}
\langle\Omega| \chi^\mu (0) | \rho(\vec{p},s) \rangle \langle \rho(\vec{p},s) |  {\chi^\nu}^{\dagger} (0) |\Omega\rangle\ + ...\ . 
\end{eqnarray*} 
Here there are contributions to the correlation function from the
$\rho^{+}$ meson, plus higher energy terms. To evaluate the correlation
functions at the hadronic level we use the formulae,
\begin{eqnarray*}
\langle\Omega|\, \chi^\mu(0) \,| \rho(\vec{p},s) \rangle  &=& \lambda \, \epsilon^{\mu}(p,s) \,  \\
\langle \rho(\vec{p},s) |\, {\chi^{\nu}}^{\dagger}(0)\, | \Omega\rangle  &=&  \bar{\lambda} \, \epsilon^{\star\nu}(p,s)\   \ ,
\end{eqnarray*}
where $\lambda$ and $\bar{\lambda}$ are the couplings of the interpolator 
to the $\rho$ at the source and sink respectively and 
%with some spin-polarisation vector $\epsilon$ and where 
$p^\mu = (E_{\rho},\vec{p})$.
We demand that the spin of the vector meson is orthogonal to its
physical momentum because the vector meson current is conserved.
The transversality condition is,
\begin{eqnarray*}
\sum_s \epsilon^{\mu} (p, s)\, \epsilon^{\star\nu} (p, s) = - \left(g^{\mu\nu} - \frac{p^\mu p^\nu}{m^2} \right) \ .
\end{eqnarray*} 
Using this relation we find that at zero momentum,
\begin{eqnarray*}
&&{G}^{00}(t,\vec{0}) = 0\ , \nonumber \\
&&{G}^{kl}(t,\vec{0}) = \delta^{kl} \lambda \, \bar{\lambda} \, e^{-{m_{\rho}t}} + ... \ .
\end{eqnarray*}

\subsection{ 3-pt Correlation Function }

\begin{figure}
\label{fig:vertex}
\begin{center}
\includegraphics[width=10cm,angle=0]{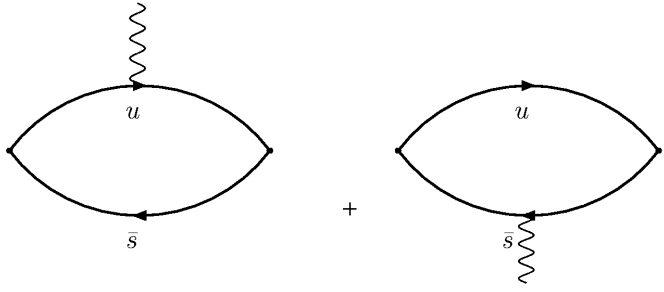} 
\end{center}
\caption{Quark flow diagrams for the $K^{*}$ meson.}
\end{figure}

The form factors are extracted from the hadronic matrix element $
\langle p',s' | J^{\alpha} | p,s \rangle$, shown diagrammatically in
Fig.~\ref{fig:vertex}.
We calculate each of the the quark sector contributions to the form
factors separately, i.e.\ in Fig.~\ref{fig:vertex} we calculate this
amplitude with the photon striking the light- and heavy-quarks
separately. The quark sector contributions are then combined to
assemble observables.  
Following Brodsky and Hiller \cite{Brodsky:1992px}, 
for the pseudo-scalar mesons, 
\begin{eqnarray*}
\left \langle p'  | J^{\alpha} | p  \right \rangle = \frac{1}{2\sqrt{E_p E_{p'}}}\, [p^{\alpha} + {p'}^{\alpha}] F_1(Q^2)
\end{eqnarray*}
%
%john - pion and kaon have only one spin state, so we omit the s and s' indices.
%
and for the vector mesons,  
\begin{eqnarray*}
\left \langle p',s' | J^{\mu} | p,s  \right \rangle &=&\
\frac{1}{2\sqrt{E_p E_{p'}}}\,\epsilon^{\prime \star }_{\alpha}(p',s') \epsilon_{\beta}(p,s)  J^{\alpha \mu \beta}(p',p) \cr
J^{\alpha \mu \beta}(p',p)  &=& - \bigg\{ G_{1}(Q^2)\, g^{\alpha \beta} \, [p^{\mu} + p'^{\mu}] +\
                                      G_2(Q^2) \, [g^{\mu \beta} q^{\alpha} - g^{\mu \alpha}  q^{\beta} ] +\
                                      G_3(Q^2) \, q^{\beta} q^{\alpha} \  \frac{p^{\mu} + p^{\prime \mu}}{2M^2} \bigg\} \ .
\end{eqnarray*}
In our calculation the mesons initially have zero 
momentum and, after scattering with the photon, one unit of momentum in the
final state. The final momentum is in the $x$-direction.
The covariant vertex functions $G_1(Q^2), G_2(Q^2)$, and $G_3(Q^2)$ are 
related to the Sachs form factors \cite{Brodsky:1992px} via,
\begin{eqnarray}
  G_Q(Q^2) &=& G_1(Q^2) - G_2(Q^2) + (1 +\frac{Q^2}{4 m^2}) G_3(Q^2) \\
  G_M(Q^2) &=& G_2(Q^2) \\
  G_C(Q^2) &=& G_1(Q^2) + \frac{2}{3} \frac{Q^2}{4 m^2} G_Q(Q^2) \, .
\end{eqnarray}
In this simulation $Q^{2} \simeq 0.22\ {\rm GeV}^{2}$.

On the lattice we define the three point function,
\begin{eqnarray*}
G^{\mu\alpha\nu}(t_2,t_1,\vec{p}\,',\vec{p}) &=& \sum_{\vec{x_1},\vec{x_2}} e^{-i \vec{p'}.(\vec{x_2} - \vec{x_1})} e^{-i \vec{p}.\vec{x_1}} \langle \Omega | \chi^\mu (x_2) J^{\alpha}(x_1) \chi^{\dagger \nu} (0) | \Omega \rangle \cr
& & \hspace{-3cm} = \sum_{i,j} \sum_{s,s'} e^{ -E_{i} (t_{2} - t_{1} )} e^{ -E_{j} t_{1} } \
     \langle \Omega | \chi^\mu  | p',s' \rangle \langle p',s' | J^{\alpha} | p,s \rangle \langle p,s  |\chi^{\dagger \nu} | \Omega \rangle \ ,
\end{eqnarray*}
where the interpolator creates a meson at the source, the current is
inserted at an intermediate time (time slice $t_{1}=14$), and finally
the state is annihilated at the sink at $t_{2}$.
We can show that the Sachs form factors can be extracted from
linear combinations of the ratios,
\begin{eqnarray*}
R^{\mu\alpha\nu}(p',p) =  \sqrt{\frac{ \left<G^{\mu\alpha\nu}(\vec{p'},\vec{p},t,t_1)\right> \, \left<G^{\nu\alpha\mu}(\vec{p},\vec{p'},t,t_1)\right>}{  \bigg{<}G^{\mu\mu}(\vec{p'},t)\bigg{>} \, \bigg{<}G^{\nu\nu}(\vec{p},t)\bigg{>}}}\ .
\end{eqnarray*}
In this notation $\mu$ is the Lorentz index of the interpolator
at the sink, $\alpha$ is the Lorentz index on the current insertion
and $\nu$ is the Lorentz index of the interpolator source.  It can be
shown that,
\begin{eqnarray*}
 G_C(Q^2) &=& \frac{2}{3} \frac{\sqrt{Em}}{E+m} \left( R^{101} + R^{202} + R^{303} \right) \\
 G_M(Q^2) &=& \frac{\sqrt{Em}}{p_x} \left( R^{133} + R^{331} \right) \\
 G_Q(Q^2) &=& \frac{m \sqrt{Em}}{p^2_x} \left( 2 R^{101} - R^{202} - R^{303} \right) \ .
\end{eqnarray*}

The charge radius is defined in terms of the charge form factor as,
\begin{eqnarray*}
\langle r^2 \rangle = -6 \frac{\partial}{\partial Q^2} G(Q^2) {\Big |}_{Q^2=0}\ .
\end{eqnarray*}
Our lattice calculations are necessarily defined at finite $Q^{2}$, but we
continue to $Q^{2}=0$ by assuming $G_C$ has a monopole form,
\begin{eqnarray*}
 G_C(Q^2) = \left ( \frac{1}{\frac{Q^2}{\Lambda^2}+1}\right)\ ,
\end{eqnarray*}
enabling the derivative to be taken. The lattice form factor calculation then defines $\Lambda$.
Motivated by the scaling of $G_{M}(Q^{2}) \over G_{C}(Q^{2})$ for
baryons at small $Q^{2}$, we extrapolate $G_{M}(Q^{2})$ to $Q^{2}=0$
by assuming,
\begin{eqnarray*}
 G_M(0) \simeq \frac{ G_M(Q^2)}{ G_C(Q^2)} \ , 
\end{eqnarray*}
for individual quark sectors.  In terms of the magnitude of the
electron charge $e$ and the mass of the meson $M$, the magnetic moment
is,
\begin{eqnarray*}
\mu_1 =   G_M (0) { e \over 2 M }\ .
\end{eqnarray*}
%The magnetic moment in units of nuclear magnetons $\mu_{N}$ is $\mu_1$
%multiplied by the ratio of the physical nucleon mass and the meson
%mass.

Our simulations are done on a large $20^3 \times 40$ lattice, $a =
0.128$ fm with 380 gauge field configurations. Our definition of the
conserved current is $O(a)$ improved and conserved \cite{Martinelli:1990ny,Leinweber:1990dv}.
%, see Martinelli
%et~al. for further details. 
We use the FLIC fermion action.
% with ape
%smear in the irrelevant terms in the conserved current for improve
%scaling. 
For further details of the simulation parameters used in this
calculation see Zanotti et~al.~\cite{Zanotti:2004dr}. We note that the
vector mesons are bound at all quark masses used in this calculation.

\section{Results}
\begin{figure}[htbp]
\begin{center}
\label{fig:rsquared}
\includegraphics[height=10cm,angle=90]{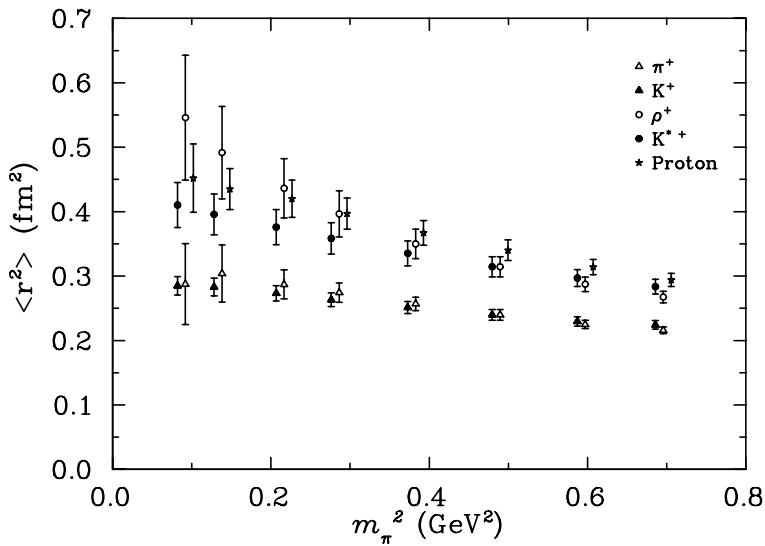} 
\end{center}
\caption{Summary of the  meson and proton charge radii.}
\end{figure}

We begin the discussion of our results with the charge radii of the
vector and pseudo-scalar mesons. From the quark model we would expect
a hyperfine interaction between the quark and anti-quark of the form
$\frac{\vec{\sigma_{q}}\cdot\vec{\sigma_{\bar{q}}}}{m_q m_{\bar{q}}}$.
The interaction is repulsive where the spins are aligned, as
in the vector mesons, and attractive where the spins are
anti-aligned, as in the pseudo-scalar mesons.  In
Fig.~\ref{fig:rsquared} we show the charge radii of the vector and
pseudo-scalar mesons.
Indeed we find that the charge radii of the vector mesons are
consistently larger than the pseudo-scalar mesons, and in fact similar
to the charge radii of the proton. 
%Nevertheless, it is fascinating that the heavier vector mesons are
%exhibiting the larger characteristic size, despite a smaller Compton
%wavelength, due to its internal structure.

\begin{figure}[tbp]
\begin{center}
\begin{tabular}{lr}
\includegraphics[height=7.5cm,angle=90]{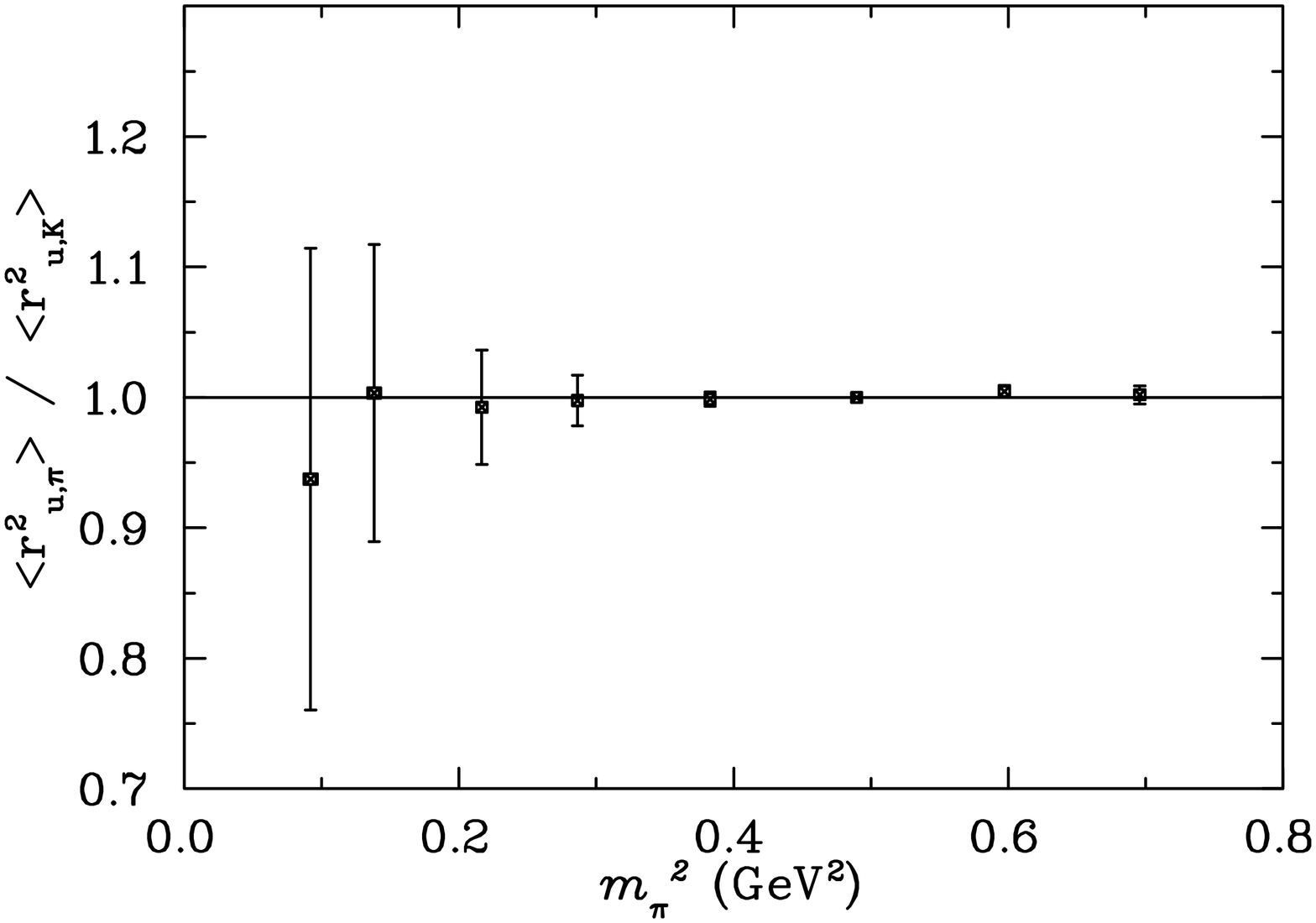} &
\includegraphics[height=7.5cm,angle=90]{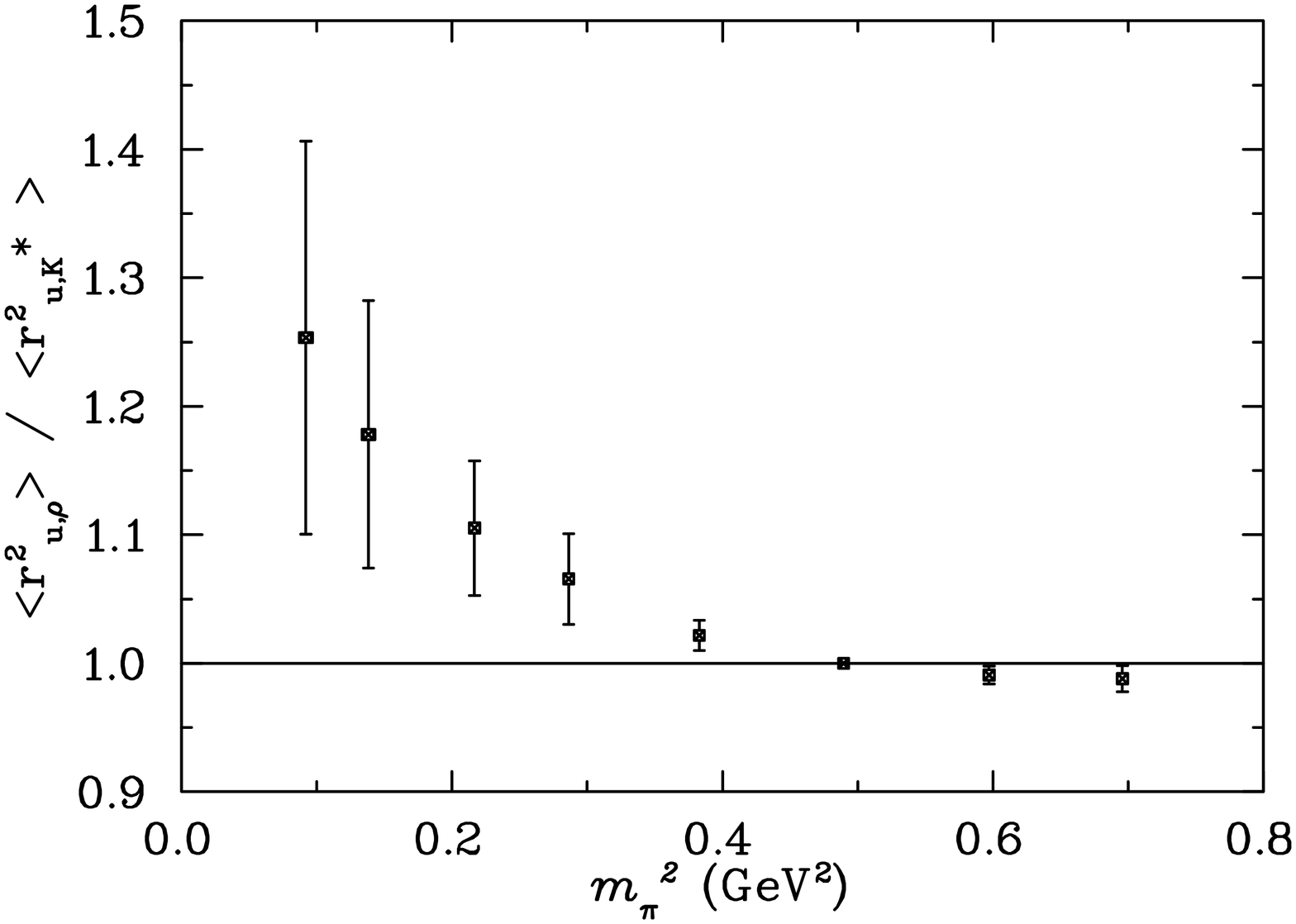}
\end{tabular}
\end{center}
\caption{\label{fig:enviro}(Left)  Ratio of the light-quark contribution to $<r^{2}>\ {\rm fm}^{2}$ in the pion and kaon.
         (Right) Ratio of the light-quark contribution to $<r^{2}>\ {\rm fm}^{2}$ in the $\rho^{+}$ 
                 and $K^{* +}$. }
\end{figure}

Next in Fig.~\ref{fig:enviro} we show the ratio of the light-quark
contribution to the charge radii of the strange and non-strange
mesons. We find that there is no evidence of the environmental
sensitivity in the light-quark contribution the pseudo-scalar mesons.
However we do find clear evidence of environmental sensitivity of the
light-quark contribution to the vector mesons at our smaller quark
masses.  The broadening of the charge distribution in the $\rho^{+}$
is consistent with the hyperfine repulsion discussed above.

Next we present our analysis of the magnetic moments of the vector
mesons. At the SU(3) flavour limit, the simple quark model predicts
that the magnetic moment of the $\rho$ meson in nuclear magnetons is
$\mu_{\rho} \simeq 1.84 \mu_{N}$, three times the magnetic moment of
the $\Lambda$ baryon. Our results are consistent with this prediction.

\begin{figure}[tbp]
\begin{center}
\begin{tabular}{lr}
\includegraphics[height=7.5cm,angle=90]{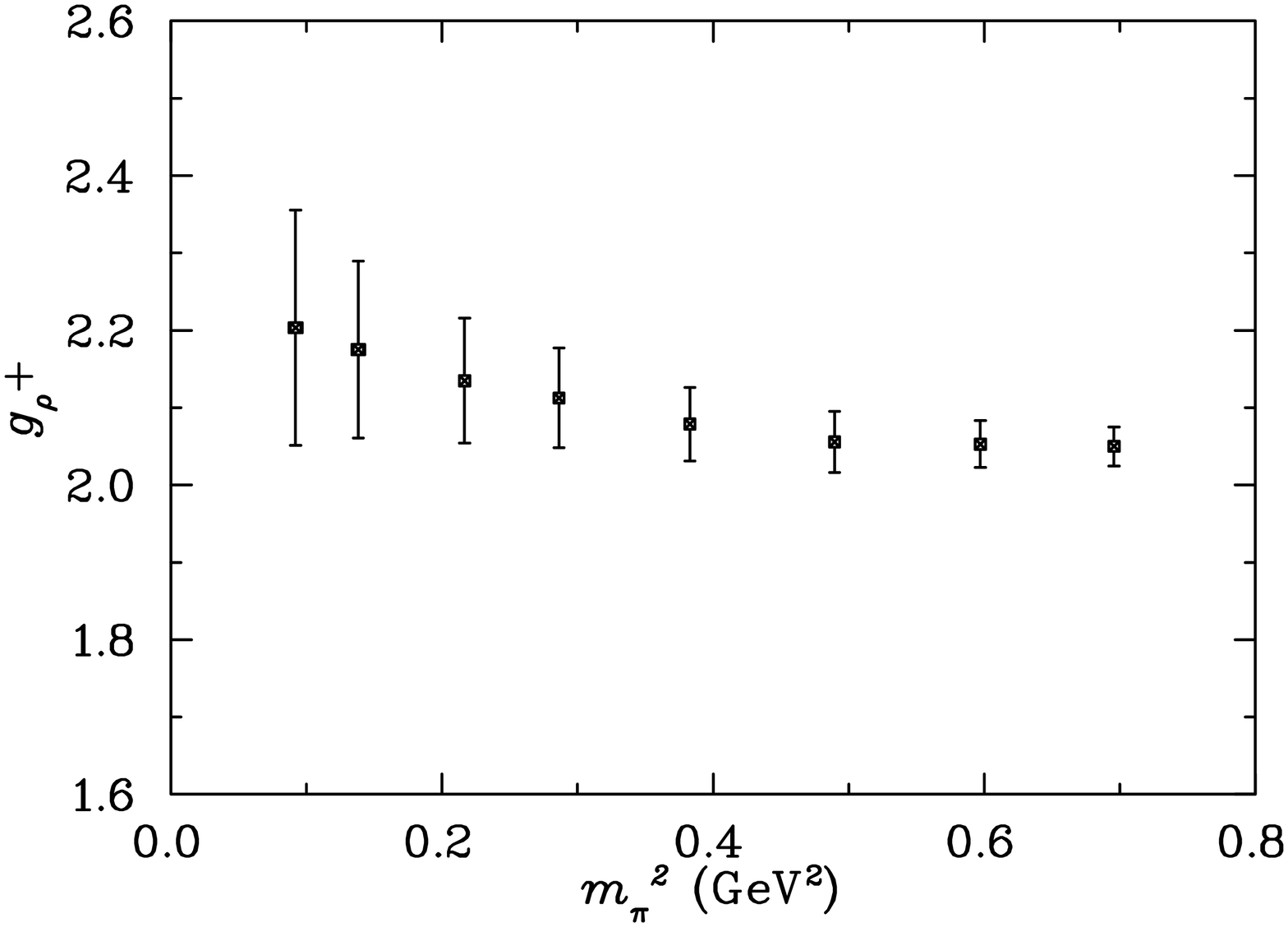} &
\includegraphics[height=7.5cm,angle=90]{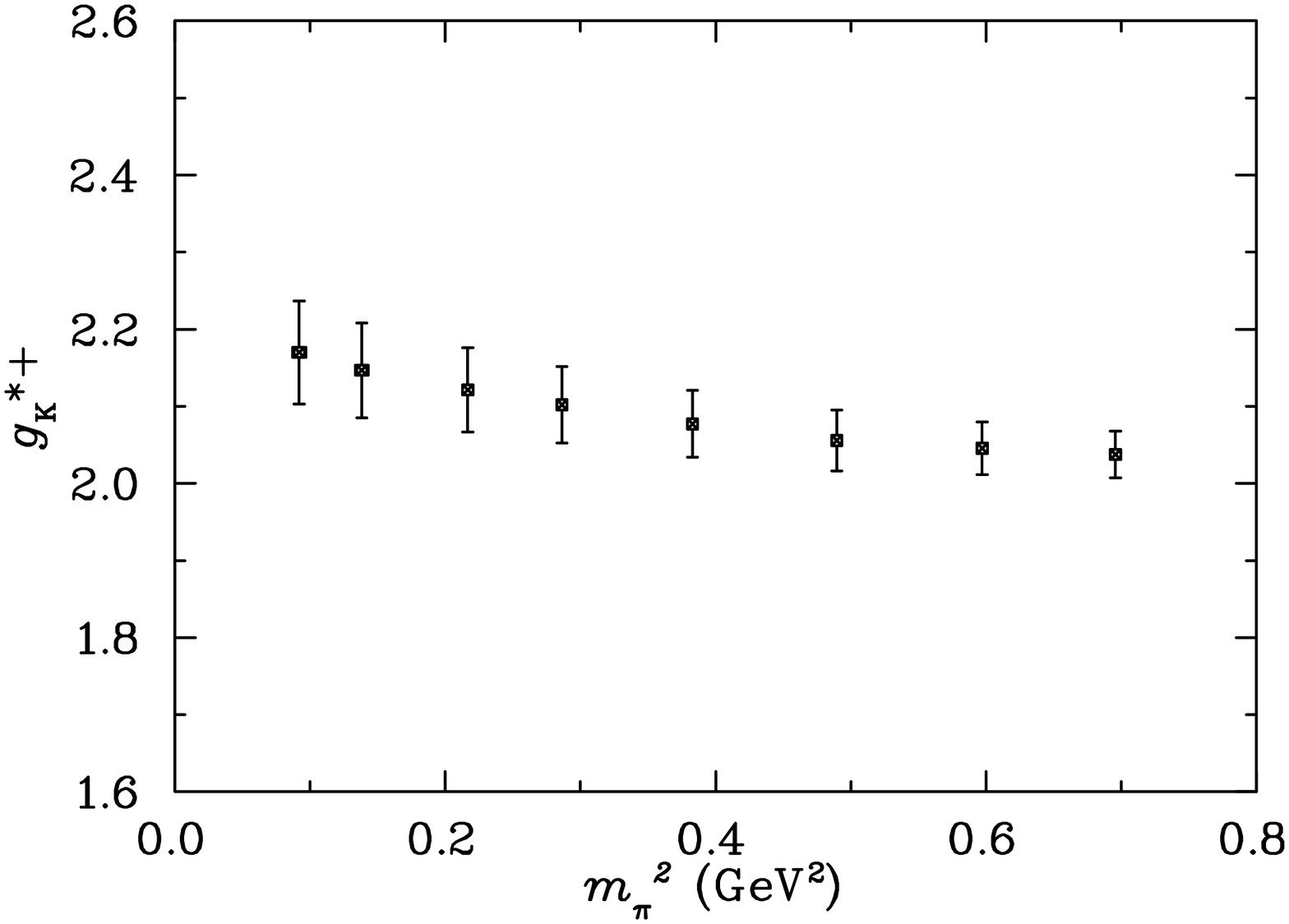}
\end{tabular}
\end{center}
\caption{
  \label{fig:gfactor}
  (Left)  G-factor of the $\rho$ meson.
  (Right) G-factor of the $K^{*}$ meson. }
\end{figure}

To easily compare with previous lattice simulations \cite{Andersen:1996qb}, in
Fig.~\ref{fig:gfactor} we report the g-factor of the vector mesons. 
%With relatively small statistical errors we find that the G-factor of
%the $\rho$ and $K^*$ mesons is consistently less than the quark model
%expectations of ~2.3. This is because in quenched lattice QCD, the
%amplitude $\rho \rightarrow \pi^{+}\omega$ is the leading non-analytic
%term a chiral effective field theory. In the real world it has a +ve
%contribution to the magnetic moment, which is not present in quenched
%QCD.  However 
We note that our calculation is consistent with the previous lattice
calculation.

\begin{figure}[tbp]
\begin{center}
\includegraphics[height=10cm,angle=90]{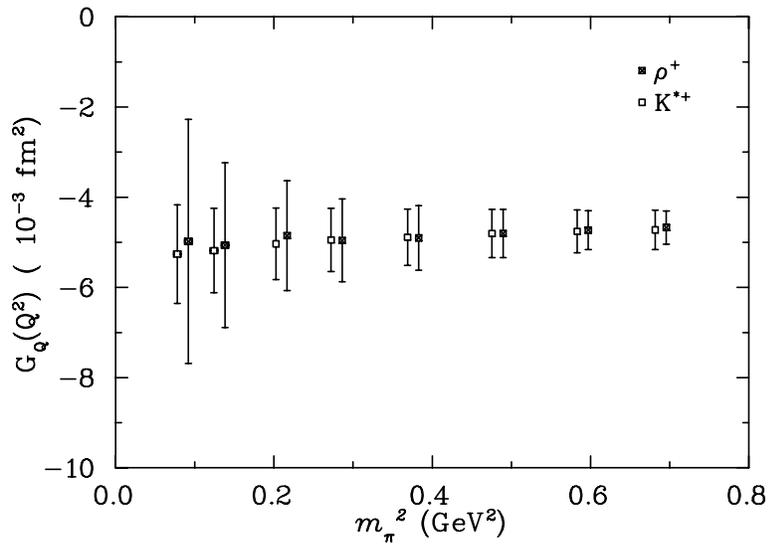} 
\end{center}
\caption{\label{fig:quadff} $\rho$-meson quadrupole form factor at
$Q^{2} \simeq 0.22\ {\rm GeV}^{2}$. }
\end{figure}

Finally in Fig.~\ref{fig:quadff} we show the quadrupole form factor of
the $\rho$. We find that the quadrupole form factor is less than zero
indicating that the spatial distribution of charge within the $\rho$ meson
is oblate.  This is in accord with the findings
of Alexandrou et~al.\cite{Alexandrou:2002nn}.

\section{Conclusions}

%In conclusion, we find that hyperfine repulsion in the vector mesons
%in quenched lattice QCD is large.  
In conclusion, we find that the charge radii of the vector mesons are
larger than the charge radii of the pseudoscalar mesons.  Indeed we
find that the trend in the charge radius of the vector mesons is
similar to the charge radius of the proton. By evaluating the quark
sector contributions to each observable separately we are able to
determine that there is significant environmental sensitivity in the
light-quark contributions to charge radii of the vector mesons.  
%We find that the G-factor of
%the $\rho$-meson is less than the quark model predictions in quenched
%lattice QCD, which we explain as a quenching artifact.  
We find that the magnetic moment of the $\rho$ is consistent with
quark model predictions at the SU(3) flavour limit, and with previous
lattice simulations.  Finally we determine that the quadrupole form
factor of the $\rho$ meson is negative which means that the
distribution of charge in the vector mesons is oblate.

\acknowledgments

We thank the Australian Partnership for Advanced Computing (APAC) and
the South Australian Partnership for Advanced Computing (SAPAC) for
generous grants of supercomputer time which have enabled this project.

%\bibliographystyle{JHEP-2} 
%\bibliography{bib}

%\begin{thebibliography}{99}

%\end{thebibliography}

\providecommand{\href}[2]{#2}\begingroup\raggedright\endgroup

\end{document}